\documentclass[twoside]{dis07}
\usepackage[latin1]{inputenc}
\usepackage[dvips]{graphicx,epsfig,color}
\usepackage{wrapfig,rotating}
\usepackage{amssymb,amsmath,array}

\begin{document}
\title{Jet Cross-Sections and $\alpha_S$ in DIS at HERA}

\author{Thomas Sch\"orner-Sadenius
%
\thanks{Talk given at DIS07 on behalf of the ZEUS collaboration.}
%
\vspace{.3cm}\\
%
Hamburg University - IExpPh \\
Luruper Chaussee 149, D-22761 Hamburg - Germany
}

\maketitle

\begin{abstract}
Measurements of inclusive-jet and dijet cross-sections in high-$Q^2$ 
deep-inelastic scattering are presented together with a short overview 
of extractions of the strong coupling parameter $\alpha_S$ 
from jets. The data samples used were
collected with the ZEUS detector at HERA-1 and HERA-2. 
The measured distributions are compared to QCD 
calculations in next-to-leading order which describe the data very well. 
The various determinations of $\alpha_S$ give a consistent picture, 
have competitive uncertainties and clearly demonstrate the running 
of the coupling predicted by QCD. 
\end{abstract}

\section{Introduction}

Measurements of jet cross-sections in high-$Q^2$ deep-inelastic scattering 
(DIS) have traditionally been used to test 
the concepts of perturbative QCD (power expansion,
factorisation, PDF universality). In addition, jet measurements in DIS 
allow precise determinations 
of the strong coupling $\alpha_S$ and are a valuable input to global fits
of the PDFs (see for example~\cite{zeus:nlofit}). 

\begin{wrapfigure}{r}{0.5\columnwidth}
\centerline{\includegraphics[width=0.45\columnwidth]{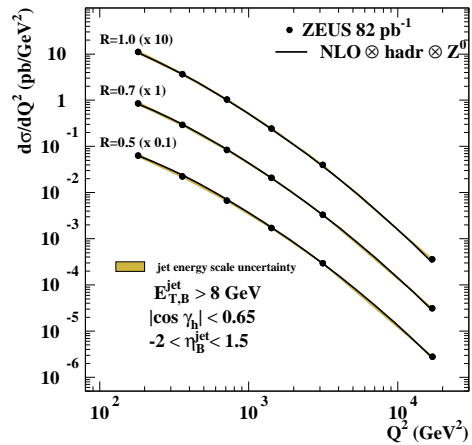}}
\caption{ZEUS inclusive-jet cross-sections d$\sigma$/d$Q^2$ for various values of $R$.}\label{fig:incl:q2}
\end{wrapfigure}

In this article measurements of jet cross-sections in high-$Q^2$ DIS are 
reported which are new in two respects: First, the first neutral-current 
(NC) jet measurement in a combined HERA-1+HERA-2 
data sample from the ZEUS experiment is presented, 
showing single- and double-differential dijet cross-sections. 
A similar measurement of such dijet cross-sections in HERA-1 data has 
recently been published by the ZEUS collaboration~\cite{zeus:dijets}. 
Second, HERA-1 
data have been used to measure inclusive jet cross-sections with different 
values of the $R$ parameter in the inclusive $k_T$ jet clustering 
algorithm~\cite{zeus:inclusive}. 
This parameter defines the distance up to which 
two particles may be joined into a new pseudo-particle in the process of 
jet clustering. An investigation of the $R$-dependence of jet cross-sections 
may prove helpful for heavy flavour physics, for hadronisation studies or for 
physics studies at the LHC. The inclusive-jet data have in addition been used 
for a new determination of $\alpha_S$. For this reason, 
this article also gives a short overview of $\alpha_S$ measurements from jets at ZEUS. 

\section{Data samples and selections}

The inclusive-jet analysis reported on was carried out in 
82~${\rm pb^{-1}}$ of data from the years 1998-2000; the dijet analysis used
in addition about 127~${\rm pb^{-1}}$ from the electron running 
period in 2004/05. Together with a ZEUS jet measurement 
in charged-current events~\cite{talks:ccjets}, this   
dijet measurement is the first jet measurement in HERA-2 data. 

\begin{wrapfigure}{l}{0.5\columnwidth}
\centerline{\includegraphics[width=0.45\columnwidth]{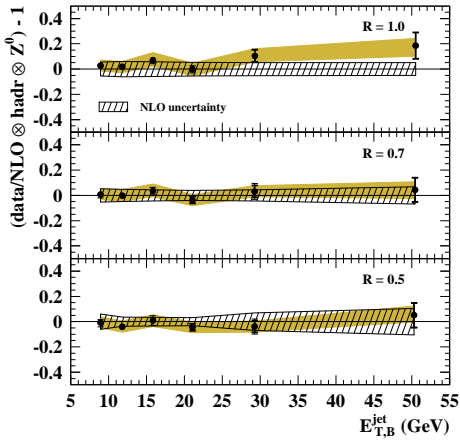}}
\caption{Ratio of data over NLO QCD for the inclusive-jet 
cross-sections d$\sigma$/d$E_T^{Breit}$ for various values of $R$.}\label{fig:incl:et}
\end{wrapfigure}

The event selections of both the inclusive-jet and dijet measurements follow 
closely that described in~\cite{zeus:dijets} and require 
high values of $Q^2 >$~125~${\rm GeV^2}$ to ensure 
relatively small theoretical uncertainties. 
Furthermore, the  
requirement -0.65~$< \cos\gamma_{had}<$~0.65 was imposed, where in 
lowest-order (Quark-Parton-Model) events $\gamma_{had}$ corresponds to 
the scattering angle of the struck parton. The cut on $\gamma_{had}$ helps 
to select phase-space regions with good acceptance and to ensure a 
good reconstruction of jets in the Breit reference frame. 

Jets were reconstructed in the Breit frame using the longitudinally 
invariant $k_T$ clustering algorithm in the inclusive mode; the 
Breit-frame pseudorapidities of the jets, $\eta^{Breit}$, were restricted to 
-2~$< \eta^{Breit} <$~1.5, and the Breit-frame transverse energies of the 
jets, $E_T^{Breit}$, were restricted to $E_T^{Breit} >$~8~GeV. For the 
dijet analysis the hardest jet was in addition required to satisfy $E_T^{Breit} >$~12~GeV. 

All data distributions were corrected for detector and QED radiation 
effects using leading-order (LO) Monte Carlo (MC) programs in a 
bin-to-bin fashion. The data were then compared to next-to-leading order 
(NLO) QCD calculations that employed either the latest PDF sets from the 
CTEQ group or the ZEUS-S PDFs. The NLO predictions were corrected 
for hadronisation effects, and in case of the inclusive-jet measurement, 
for $Z^0$ contributions.  

\section{Experimental and theoretical uncertainties}

The experimental uncertainties are dominated by the uncertainty in the
jet energy scale which is assumed to be 1-3~$\%$, depending on the jet
$E_T$. Resulting uncertainties on the measured cross-sections are typically
5-10~$\%$. The next-largest uncertainty stems from the model uncertainty in
the unfolding of the measured distributions to the hadron level; further
sources of uncertainty like the effects of selection cut variations are
typically much smaller.

On the theoretical side, the effect of higher orders not considered in the
perturbative expansion and the uncertainties on the input PDFs give the
largest contributions. The former is typically estimated by variations of the
renormalization scale $\mu_R$ by an arbitrary, but customary amount; the
effects on the cross-sections are typically in the order of 5-10~$\%$ for the
inclusive jet measurements and slightly larger for the dijet measurements. The
effect of the PDFs is somewhat smaller, depending on the region of phase-space
considered. The effects of the uncertainties on $\alpha_S$, on the
hadronization corrections and on the factorization scale are much smaller. It
should be noted that in almost all experimental bins the theoretical
uncertainty is significantly larger than the experimental one.

\section{Inclusive-jet cross-sections}

Inclusive-jet cross-sections at high $Q^2$ were measured as functions of 
the jet transverse energy in the Breit frame, $E_T^{Breit}$ and of $Q^2$ for 
three different values of $R$: 0.5, 0.7, 1.0 (it turns out that for higher 
(lower) values of $R$ the uncertainties due to missing higher 
orders (hadronisation effects) become drastically larger). 

\begin{wrapfigure}{r}{0.5\columnwidth}
\centerline{\includegraphics[width=0.45\columnwidth]{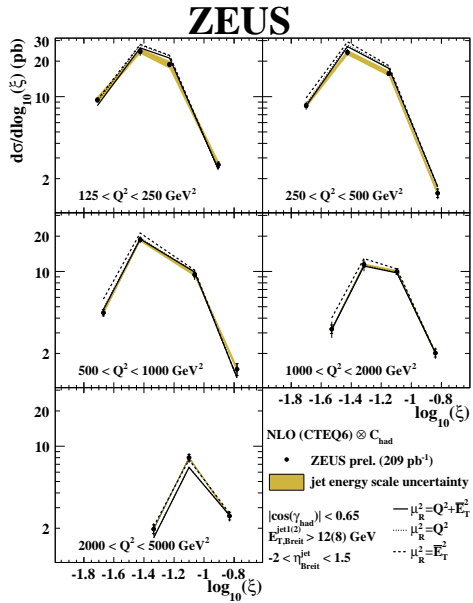}}
\caption{ZEUS dijet cross-sections d$\sigma$/d$\log\xi$ in different regions of $Q^2$.}\label{fig:dijets}
\end{wrapfigure}

The resulting
cross-section as a function of $Q^2$ is shown in Figure~\ref{fig:incl:q2}. The
data (points) are compared to
the NLO QCD prediction (lines); the description of the data by theory is
good. Figure~\ref{fig:incl:et} shows the ratio between the measured
cross-sections as functions of jet transverse energy $E_T$ and the NLO
calculations. The measurement can be seen to be dominated by the
(correlated) jet energy-scale and theoretical errors 
(grey and hashed areas, respectively),
except for the highest-$E_T$ (and also highest-$Q^2$) points for which the
statistical uncertainties are large (10~$\%$). Within all errors, the data are
well described by the theoretical predictions. The data have also been plotted as
a function of the jet radius parameter $R$, integrated over all $E_T$ and $Q^2$
values (not shown). A more or less linear increase of the cross-section with
$R$ can be observed, which is to be expected since with larger
radii the jet algorithm picks up more and more of the jet's energy.

\section{Dijet cross-sections}

Figure~\ref{fig:dijets} shows the dijet cross-section for jets above 12 and 
8~GeV, respectively, as a function of $\log_{10}\xi$ in different regions of
$Q^2$. In leading order, the observable $\xi$
corresponds to the momentum fraction carried away from the incoming proton by
the struck parton. These cross-sections therefore depend on the two variables
relevant for the PDFs, the energy scale and the
momentum fraction, and thus might be useful for an improvement of
the PDF precision. 

The data are again compared to NLO QCD predictions, using various squared
renormalization scales, namely $\mu_R^2 = Q^2 + \overline{E_T}^2$, $\mu_R^2 =
Q^2$ and $\mu_R^2 = \overline{E_T}^2$. The shaded band indicates again the jet
energy-scale uncertainty that is assumed to be correlated from bin to
bin. Taking into account all uncertainties, the data are well described by the
NLO predictions.  However, it should be noted that in different regions of
phase-space different scale choices are required to achieve this good
agreement: The low-$Q^2$ data points are better desribed using the squared
scale $\mu_R^2 = Q^2+\overline{E_T}^2$, whereas at high $Q^2$
$\mu_R^2 = \overline{E_T}^2$ seems more appropriate.   

\begin{wrapfigure}{l}{0.5\columnwidth}
\centerline{\includegraphics[width=0.45\columnwidth]{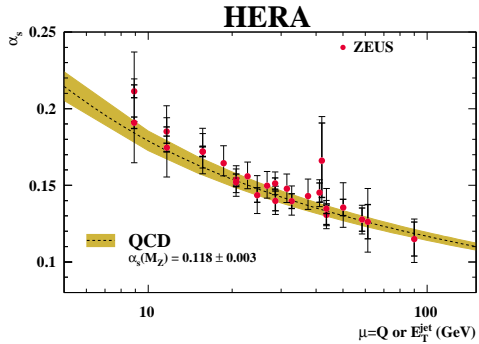}}
\caption{Comparison of various ZEUS $\alpha_S$ measurements from jets in DIS with 
the QCD prediction}\label{fig:alphas}
\end{wrapfigure}

\section{Determining $\alpha_S$ with jets at ZEUS}

The strong coupling parameter $\alpha_S$ has been determined from a variety of
QCD measurements at ZEUS, the latest result coming from the inclusive-jet 
measurements described above~\cite{zeus:inclusive}: 
 $\alpha_S = 0.1207 \pm 0.0014(stat.) \pm ^{0.0035}_{0.0033}(exp.) \pm ^{0.0022}_{0.0023}(th.)$
The various measurements at different energy scales
are well described by the behaviour of the coupling as expected from QCD, see
Fig.~\ref{fig:alphas}. A combination of ZEUS and H1 $\alpha_S$
measurements has been carried out recently~\cite{claudia.alphas} and has led
to a value of $\alpha_S = 0.1186 \pm 0.0011(exp.) \pm 0.0050(th.)$. The good
agreement of the various measurements with each other and with the world
average value indicates the high level of our present understanding of
QCD. Nevertheless, it should be pointed out that the $\alpha_S$ measurements
from jets in DIS suffer from large theoretical uncertainties which are mostly
due to missing higher orders in the perturbative expansion of the presently
available QCD predictions.   

\section{Conclusions}

With the advent of HERA-2 analyses and the possibility of combined
HERA-1/HERA-2 analyses, QCD studies with jets at HERA enter a new
regime. In this contribution, measurements of
inclusive-jet and dijet cross-sections have been discussed together with
extractions of the strong coupling parameter $\alpha_S$ from jet and other DIS
measurements. Although many jet measurements are limited by theoretical
uncertainties, the impact of further jet
measurements on our knowlegde of PDFs and $\alpha_S$ should be assessed and
exploited. 


\begin{footnotesize}


\end{footnotesize}


\end{document}